\documentclass[12pt,letterpaper]{article}
\usepackage{osajnl}

\newcommand{\degree}{\ensuremath{^\circ}}
\begin{document}

\OSAJNLtitle{Optical Design of the Atacama Cosmology Telescope and
  the Millimeter Bolometric Array Camera}

\newcommand{\princeton}{$^1$}
\newcommand{\penn}{$^2$}
\newcommand{\cardiff}{$^3$}
\newcommand{\ubc}{$^4$}
\newcommand{\umn}{$^5$}
\newcommand{\puastro}{$^6$}

\OSAJNLauthor{
  J. W. Fowler\princeton,
  M. D. Niemack\princeton, 
  S. R. Dicker\penn, 
  A. M. Aboobaker\princeton$^,$\umn, 
  P. A. R. Ade\cardiff,
  E. S. Battistelli\ubc,
  M. J. Devlin\penn, 
  R. P. Fisher\princeton,
  M.    Halpern\ubc,
  P. C. Hargrave\cardiff,
  A. D. Hincks\princeton,
  M.    Kaul\penn,
  J.    Klein\penn, 
  J. M. Lau\princeton,
  M.    Limon\penn,
  T. A. Marriage\princeton$^,$\puastro, 
  P. D. Mauskopf\cardiff,
  L.    Page\princeton, 
  S. T. Staggs\princeton,
  D. S. Swetz\penn, 
  E. R. Switzer\princeton,
  R. J. Thornton\penn, 
  C. E. Tucker\cardiff
}

%

\address {\princeton\  
  Princeton University Department of Physics, \\
  Jadwin Hall, Washington Road, Princeton, New Jersey 08544, USA}
\address {\penn\  
  University of Pennsylvania Department of Physics and Astronomy, \\
  David Rittenhouse Laboratory, 209 South 33rd Street,
  Philadelphia, Pennsylvania 19104, USA}
\address {\cardiff\  
  Cardiff University
  School of Physics and Astronomy, \\
  Queens Buildings,
  5 The Parade,
  Cardiff CF24 3AA, Wales, United Kingdom}
\address {\ubc\  
  University of British Columbia
  Department of Physics and Astronomy, \\
  6224 Agricultural Road, 
  Vancouver, British Columbia V6T 1Z1, Canada}
\address{\umn\ 
  Present Address: University of
  Minnesota School of Physics and Astronomy, \\ 
  116 Church St. SE,
  Minneapolis, Minnesota 55455, USA}
\address {\puastro\  
  Present Address: Princeton University 
  Department of Astrophysical Sciences, \\
  Peyton Hall, Ivy Lane, Princeton, New Jersey 08544, USA}

\OSAJNLemail{jfowler@princeton.edu}

\begin{abstract}
  The Atacama Cosmology Telescope is a 6-meter telescope designed to
  map the Cosmic Microwave Background simultaneously at 145\,GHz,
  215\,GHz, and 280\,GHz with arcminute resolution.  Each frequency
  will have a 32 by 32 element focal plane array of TES bolometers.
  This paper describes the design of the telescope and the cold
  reimaging optics, which is optimized for millimeter-wave
  observations with these sensitive detectors.
\end{abstract}



\section{Introduction}
The Atacama Cosmology Telescope (ACT)\cite{Fowler2004} will observe
the oldest light in the universe, the cosmic microwave background
(CMB), mapping it on arcminute to degree scales.  The CMB is relic
thermal radiation released when the early universe had cooled enough
for the primordial plasma to form a neutral gas, allowing light to
stream freely ever since.  The CMB's blackbody spectrum has redshifted
with the expansion of the universe to a present temperature of
2.7\,K\@.  The temperature is uniform to tens of parts per million.
Recently, the Wilkinson Microwave Anisotropy Probe
(WMAP)\cite{Bennett2003ApJS} has measured the CMB power spectrum (the
amplitude of temperature fluctuations in the CMB as a function of
angular scale) at resolutions as fine as 0.3\degree. The data from
WMAP in conjunction with other experiments \cite{Kuo2004,Leitch2005,
Jones2006, ODwyer2005,Readhead2004mosaic }---some at higher
resolution---permit estimates of the universe's global curvature and
other properties with unprecedented precision.\cite{Spergel2003}

Maps of the CMB with high resolution from ACT---combined with optical,
ultraviolet and X-ray measurements---will further constrain
inflationary models of the early universe, constrain the equation of
state of dark energy in the universe, probe light neutrino masses down
to $m_\nu\simeq$ 0.1\,eV, and map the mass distribution of the
universe.\cite{Kosowsky2003,Verde2002,Hernandez2006} Such science will
require measurements of the CMB temperature in multiple frequency
bands near the null in the Sunyaev-Zel'ovich (SZ)
effect\cite{Carlstrom2002} spectrum (217\,GHz) to a precision of a few
microkelvin at resolutions approaching one arcminute.
Arcminute resolution at these frequencies requires telescopes in the 5
to 10 meter range.  As a compromise between cost and angular
resolution, ACT has a 6-meter (projected diameter) primary mirror. In
this paper we outline the optical design that we have developed to
meet our science requirements.

Around 217\,GHz, atmospheric absorption permits observations only at
high, dry sites. The Atacama plateau in northern Chile offers an
excellent combination of observing conditions, sky coverage, and
accessibility and was selected as the site for ACT\@.  The telescope
described here has been built, and it has been shipped to Chile from
its construction site in Port Coquitlam, British Columbia.



\section{Telescope and camera overview
  \label{sec:tel_overview}}

\begin{table}[tbh]
\caption{\label{requirements}Requirements and features of the ACT optics.}
\begin{tabular}{l}\hline
 Warm Telescope Optics\\ \hline
$\bullet$\ Clear aperture (off-axis optics) to minimize scattering and
    blockage. \\
$\bullet$\ 6-meter primary mirror and  2-meter (maximum)
    secondary mirror diameters. \\
$\bullet$\ Very fast primary focus ($F \le 1$) to keep the telescope
    compact.\\
$\bullet$\ Large ($1.0\degree$) and fast ($F \sim 2.5$)
    diffraction-limited focal plane. \\
$\bullet$\  Ground loading (due to spillover) smaller than
    atmospheric loading. \\
$\bullet$\  Space for structure and cryogenics between primary mirror
    and Gregorian focus. \\
$\bullet$\ Entire telescope must scan several degrees in azimuth at 10
    cycles per minute.\\

\\ \hline
{ Cold Reimaging Optics for MBAC}\\ \hline
$\bullet$\  Bandpasses 20--30\,GHz wide, centered near 145,
215, and 280\,GHz. \\
$\bullet$\  Approximately $22'$ square field of view in each band.\\
$\bullet$\ Diffraction-limited resolution on three 34\,mm by
    36\,mm arrays.\\
$\bullet$\  Well-defined Lyot stop in all bands
    to maximize illumination of the primary.\\
$\bullet$\ Ghost images due to stray light no brighter than the
    diffraction-limited sidelobes.\\
\end{tabular}
\end{table}

Meeting the ACT science goals requires extreme sensitivity, better
than ten microkelvin rms uncertainty in map pixels of three square
arcminutes.  Even with sensitive modern millimeter-wave detectors,
large focal planes containing many hundreds of detectors, months of
integration time, and careful control of systematics are all
essential.  This section discusses the major requirements and features
of our approach, which are summarized in Table~\ref{requirements}.

The fundamental requirement of the ACT and MBAC optics design is that
the telescope and camera must reimage the sky onto a focal plane filled
with detectors $\sim1.0$\,mm in size, and that the image be
diffraction-limited.  The design is subject to geometric
limitations on the size and separation of the mirrors.  Control of
stray light is also of particular importance, since the ACT detectors
will be used without feedhorns.  Spillover radiation from the ground
around the telescope must be prevented from reaching the detectors,
and reflections and scattering within the optics must be minimized.

CMB experiments deliberately modulate their sensitivity to cosmic
signals in order to reduce the impact of drifts in detector response,
such as $1/f$ noise.  Typical modulation methods involve using an
optical chopping mirror or scanning the entire telescope in azimuth.
In either case, the telescope beam moves rapidly back-and-forth on
timescales faster than the $1/f$ knee of any low-frequency noise.  We
chose the scanning method, because it avoids a chopping flat's most
intractable scan-synchronous variations, including primary beam shape,
ground pickup pattern, and mirror emission.  Our normal observing mode
will be to scan the 50-ton telescope in azimuth over a 5\degree\ range
repeating every 5 to 6 seconds, while holding the elevation fixed
(typically at 48\degree).  This motion places considerable rigidity
requirements on the telescope structure (see
Section~\ref{sec:tolerancing}).  Observing at fixed elevation
ensures that the large gradient in atmospheric emission enters the
camera as a constant addition, not as an AC term synchronized with the
signal.  To maintain a constant speed for as much of the scan as
possible, we have aimed for brief acceleration periods of 300\,ms at
either end of each scan.

ACT will make simultaneous observations at 145, 215, and 280\,GHz to
distinguish variations in the primordial CMB from secondary
anisotropies such as SZ galaxy clusters and foregrounds such as
galactic dust and point sources.\cite{Huffenberger2005}
ACT's receiver, the Millimeter Bolometric Array Camera (MBAC) will
contain a 32$\times$32 array of transition edge sensor (TES)
bolometers\cite{Benford2003,Marriage2006} at each of the three
frequencies. The arrays will be cooled to 0.3\,K by a closed-cycle
helium-3 refrigeration system\cite{Devlin2004Cryo,Lau2006Cryo}.
Because the TES detectors are bolometric, the ACT optics must also
have optical filters to define the bandpass for each camera.

The ACT detectors are 1.05\,mm square and are spaced on a 1.05\,mm
(horizontal) by 1.15\,mm (vertical) grid.  Detectors aimed less than
half a beamwidth apart on the sky fully sample the field of view in a
single pointing.  This is advantageous for minimizing detector and
atmospheric noise in mapmaking\cite{Griffin2002}, and ACT detectors
reach this ideal at the lower frequencies.  The effective focal
length is 5.2\,m for all arrays, giving a detector spacing of $44''$
(horizontal) and $48''$ (vertical) on the sky.  This spacing is half
the expected beam size at 145\,GHz.
The use of a fast final focus at the detectors will allow MBAC to map the
sky rapidly without compromising diffraction-limited imaging
performance.


\section{Gregorian telescope optics
  \label{sec:warm_opt}}

The two-reflector Atacama Cosmology Telescope was optimized to have
the best possible average performance across a square-degree field of
view by varying the mirror shapes, angles, and separation.  This
compromise balances the various classical telescope aberrations for
point images against each other.  The design process for ACT used both
analytic and numerical methods.  Numerical methods alone might seem
sufficient, because the end result of a global optimization is
independent of the starting design.  But the telescope parameter space
is large and complicated, and we found it critical to enter the
numerical stage with a good analytic design.  We used the Code~V
optical design software\cite{OpticalRes} to optimize the telescope
design and to analyze its performance.

Our initial analytic designs met the Dragone condition
\cite{Dragone1982,Dragone1983} to minimize astigmatism, following the
implementation of Brown and Prata\cite{Brown94}.  This condition also
minimizes geometrical cross-polarization.\cite{Mizusawa} A comparison
of Gregorian and Cassegrain solutions showed that in otherwise
equivalent systems, the Gregorian offered more vertical clearance
between the secondary focus and the rays traveling from the primary to
the secondary mirror.  The extra clearance leaves more space for our
$\sim$1\,m$^3$ cryostat, so the Gregorian was chosen for ACT\@.

A simple Gregorian telescope satisfying the Dragone condition did not
meet the diffraction-limited field of view requirement, but it was
taken as the starting point for the numerical stage. The system was
optimized by minimizing the rms transverse ray aberration at field
points across the focal plane.  Six design parameters were allowed to
vary: the two conic constants, the relative tilt of the primary and
secondary axes, the secondary radius of curvature, and the location
and tilt of the Gregorian focal plane.  The primary focal length was fixed
at exactly 5\,m to keep the telescope compact.  We found that requiring
the primary and secondary mirrors to be coaxial did not substantially
degrade image quality, so we imposed this constraint to simplify
manufacturing and alignment of the telescope.

Our final design approximates an ideal aplanatic Gregorian telescope,
a system with no leading-order spherical aberration or coma in the
focal plane.\cite{Schroeder2000} Strehl ratios $S$ were estimated by
calculating $\sigma$, the rms optical path variation over a large
bundle of rays, and taking \cite{BornWolf7} $\ln S\approx -(2 \pi
\sigma)^2$.  Over a 1.0\degree\ square field at the Gregorian focus,
the Strehl ratio everywhere  exceeds 0.9 at 280\,GHz.

\begin{table}[tbh]
  \begin{center}
  \caption{\label{table_tele_shape}
    Atacama Cosmology Telescope mirror shapes.$^a$}
  \begin{tabular}{lrrrrrr}\\\hline
    Mirror & $z_\mathrm{vert}$\,(m) & $R$\,(m) & $K$ & $y_0$\,(m) &
    $a$\,(m) & $b$\,(m) \\\hline 
    Primary & $0.0000$ & $-10.0000$ &
    $-0.940935$ & $5.000$ & $3.000$ & $3.000$ \\ 
    Secondary & $-6.6625$ &
    $2.4938$ & $-0.322366$ & $-1.488$ & $1.020$ & $0.905$ \\ 
    Gregorian focus$^b$ & $-1.6758$\\\hline
  \end{tabular}
  \end{center}

\hspace{.5in} $^a$ Equation~\ref{eq:mirrors} gives the full shapes;
parameters and axes are defined in the text.

\hspace{.5in} $^b$ Best-fit focal plane location for objects at
infinity.
\end{table}

\begin{figure}[tbh]
  \centering
  \includegraphics[width=10cm]{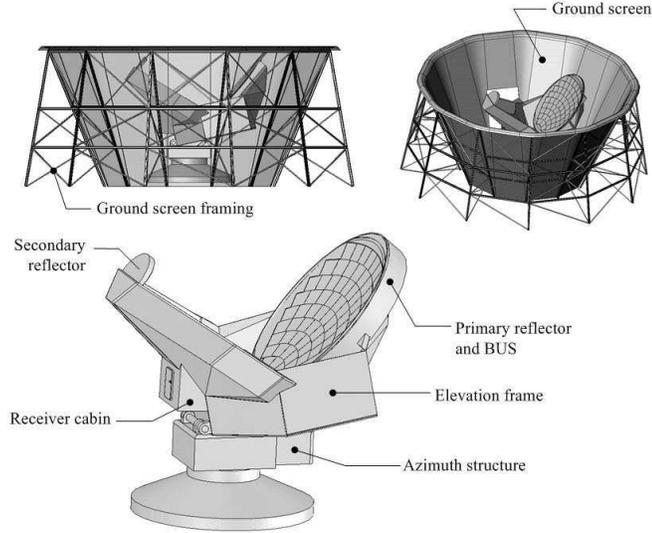}

  \caption{\label{fig:telescope} 
    The ACT telescope.  The mechanical design has a low profile; the
    surrounding ground screen completely shields the telescope from
    ground emission.  The screen also acts as a weather shield.  An
    additional ground screen (not shown) mounted on the telescope
    hides the secondary and half the primary from the vantage point of
    the lower diagram.  This inner ground screen is aluminum painted
    white to reduce solar heating.  The primary mirror is $\sim$7\,m
    in diameter including its surrounding guard ring.  ``BUS'' refers to
    the mirror's back-up structure.  (Figure credit: AMEC Dynamic
    Structures) }
\end{figure}

\begin{figure}[tbp]
  \centering
  \includegraphics[width=\textwidth]{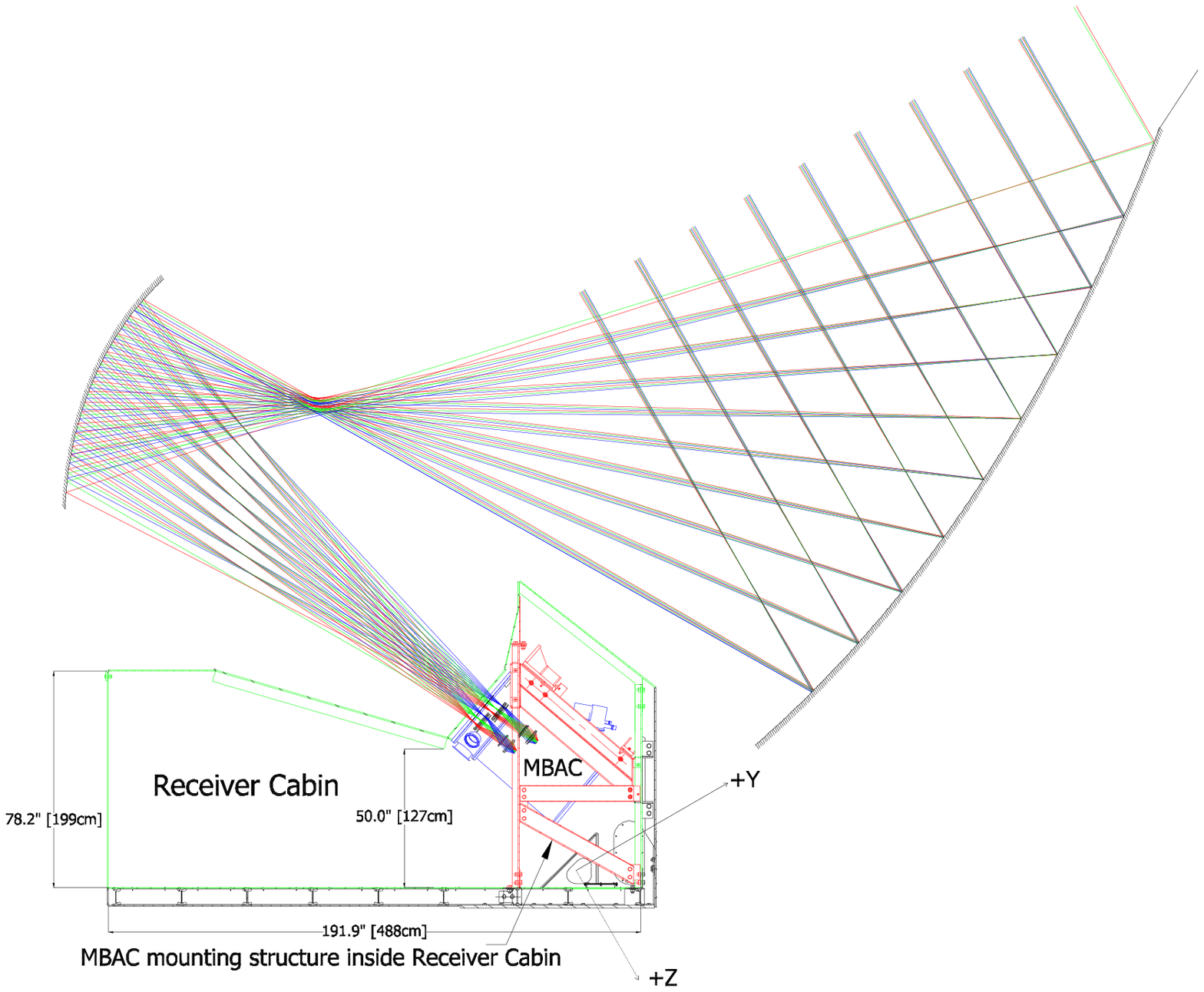}
  \caption{\label{fig:rays}
    (Color online)
    Rays traced into the MBAC cryostat (mounted at the far right of
    the receiver cabin). The stowed position is shown, corresponding
    to an elevation of 60\degree\ (generally, observations will be in
    the 40\degree--50\degree\ range).  The rays are traced from the
    central, highest, and lowest fields in the 280\,GHz camera (higher
    in the cryostat) and the 215\,GHz camera.  Both the 215\,GHz
    camera and the 145\,GHz camera (not shown) lie outside the $x=0$
    midplane, relieving any apparent conflicts between filters and
    lenses from different cameras.  The figure also shows the size and
    shape of the ACT receiver cabin, as well as the coordinate axes of
    Equation~\ref{eq:mirrors}.}
\end{figure}

The two mirrors are off-axis segments of ellipsoids in the final ACT
design.  Figure~\ref{fig:telescope} contains mechanical drawings, while
Figure~\ref{fig:rays} presents a ray trace and shows the $z$ and $y$
axes.  The parameters of each mirror are listed in
Table~\ref{table_tele_shape}.  Both shapes can be described by
\begin{equation}
\label{eq:mirrors}
z(x,y) = z_\mathrm{vert} +
\frac{(x^2+y^2)/R}{1+\sqrt{1-(1+K)(x^2+y^2)/R^2}},
\end{equation}
where $z$ is along the shared axis of symmetry (see axes on
Figure~\ref{fig:rays}), $z_\mathrm{vert}$ is the vertex position
(the primary vertex defines $z$=0), $R$ is the radius of
curvature at the vertex, the conic constant $K = -e^2$, and $e$ is the
ellipsoid eccentricity.  The usable region of each mirror is bounded
by an elliptical perimeter.  When projected into the $xy$ plane, these
boundaries are centered at $(x,y)=(0,y_0)$ and have semi-major and
semi-minor axes of $a$ and $b$ in the $x$ and $y$ directions,
respectively.  The primary projection is circular, with $a=b$.

Diffraction at the small aperture stop (in the cryogenic camera) can
lead to systematic errors, particularly if it loads the detectors with
radiation emitted by ambient-temperature structures near the two
mirrors.  To minimize this spillover effect, each mirror is surrounded
by a reflective aluminum ``guard ring.''  The rings enlarge the mirror
area beyond the geometric image of the aperture stop; they ensure that
most radiation reaching the detectors comes from the cold sky, in
spite of diffraction at the cold stop.
%
%

The ACT design also ensures that there is at least one meter of
clearance between any ray approaching the secondary and the top of the
Gregorian focal plane used by MBAC\@. The clearance allows room for a
receiver cabin that will protect the cryostat and its supporting
electronics from the harsh environment of the Atacama desert.

AMEC Dynamic Structures has designed, modeled, and built the
telescope's mechanical structure.\cite{AMEC} KUKA Robotics provided
motion control.\cite{KUKA} The primary mirror and secondary surfaces
consist of 71 and 11 aluminum panels, respectively.  Forcier Machine
Design \cite{Forcier} produced all of the panels.  The panels were
surveyed one at a time by a coordinate measuring machine and were
found to have a typical rms deviation from their nominal shapes of
only 2--3\,$\mu$m.  We measure the positions of all the panel surfaces
relative to telescope fiducial points with a Faro laser tracking
system.\cite{Faro} Four manually adjustable screw-mounts on the back
of each panel then permit precise repositioning.  To date, we have
aligned the secondary panels to approximately $15\,\mu$m rms.  A
subset of twelve primary panels has also been aligned to $30\,\mu$m
rms and monitored in detail over two twenty-four hour periods.  This
subset includes a cluster of six contiguous panels near the edge of
the primary and six others distributed uniformly over the remaining
area.  We find that the primary expands thermally as if it were a
single aluminum structure, except for mirror panels directly
illuminated by the sun.  We intend to adjust the mirror facets
annually, if necessary.

Considerable effort has gone into ensuring the best possible
performance for the azimuthal scanning of the telescope, a difficult
task given its size and weight ($\sim$50 tons).  The telescope meets
the scanning target of $\pm 2.5\degree$ at an angular speed of 2.0\degree/s
with a turn-around time of 300\,ms. Encoders mounted on the azimuthal
and elevation axes give 27-bit readings of the telescope orientation.
We have found that the pointing during scans is repeatable to better
than $4''$. Turn-arounds cause vibrations in the structure which
induce brief $\sim$$8''$ ``shuddering'' movements in elevation and an
unavoidable $\sim$$10''$ ``bounce'' in azimuth, due to the finite
bandwidth of the drive servos.  The azimuth bearing is driven by a
pair of counter-torquing helical gears, eliminating backlash.  The
elevation errors occur only during the accelerations, while the
azimuth bounce damps out exponentially $\tau\approx200$\,ms after the
acceleration ends.  Studying a large number of successive scans has
shown that the rms deviation from the average scan shape is no
more than $6.5''$ with 400\,Hz sampling, demonstrating good
repeatability of the scan pattern.

ACT's compact design minimizes accelerations of the secondary and
especially the cryogenics (which are near the rotation axis),
simplifying mechanical design and helping to maintain refrigerator
stability. The fast Gregorian focus ($F\sim$2.5) keeps the vacuum
window for the detector cryostat from being too large.
Figure~\ref{fig:rays} shows the size and shape of the receiver cabin.

Actuators can move the secondary mirror structure in the $y$ and $z$
directions by $\pm 1$\,cm from the nominal position and can tilt it up
to $\pm1\degree$ in elevation or azimuth.  We anticipate having to
refocus in response to changes in ambient temperature or observing
angle.  We plan to operate the actuators as infrequently as possible,
consistent with holding the primary-secondary
distance to within $\pm100\,\mu$m of nominal.


\section{Cold reimaging optics in MBAC
  \label{sec:cold_opt}}

Many possible architectures for the cold optics were studied,
including all-reflecting designs, all-refracting designs, and hybrids
of the two.  We also compared designs of a single camera having
dichroic filters to segregate the frequencies
against a three-in-one camera design using a separate set of optics
for each frequency.  The final MBAC design uses only refractive optics
instead of mirrors and employs the three-in-one approach.

\subsection{MBAC architecture}
Off-axis reimaging mirrors were studied by combining the equivalent
paraboloid approximation\cite{Rusch90} with the Dragone
condition\cite{Brown94}, then explored through numerical
optimization. They were rejected because the twin demands of image
quality and a wide field-of-view led to designs too large to fit in a
cubic-meter cryostat.  For off-axis mirrors, the compromises between
image quality and access to a cold image of the primary were also
unacceptable. On-axis mirrors violated the requirement of an
unobstructed aperture.

We have built dichroic beamsplitting filters as large as 15\,cm
diameter and metal mesh filters up to 30\,cm
diameter.\cite{Ade2006} Dichroics reflect one band and therefore must
be flat to $\sim\lambda/40 \approx 25\,\mu$m at 280\,GHz.  Our optical
designs required dichroics larger than any so far produced, and we
considered their production and mounting too great a risk.

We chose a camera architecture with a separate set of cold lenses for
each frequency, eschewing both cold mirrors and dichroic
beamsplitters.  There are several advantages of this design:
anti-reflection coatings and capacitive mesh filters generally have
higher transmission---and are easier to optimize---for narrow bands;
the mechanical design is simpler, more compact, and easier to align;
and the three cameras are modular and can be removed from the cryostat
separately for easy maintenance or for deploying MBAC in
stages.  The disadvantage is that each camera observes a different
area of sky.  Maps made with separated cameras do not completely
overlap, though ACT's observing plans mitigate the problem.  ACT's
scanning motion (Section~\ref{sec:tel_overview}) ensures that the
215\,GHz and 145\,GHz cameras observe most of the same sky region in a
single scan, and rotation of the earth moves fields on the sky from
MBAC's lower-elevation cameras to the upper one (or vice-versa) in
less than 15 minutes.

\begin{figure}[thb]
  \centering
  \includegraphics[width=3.5in]{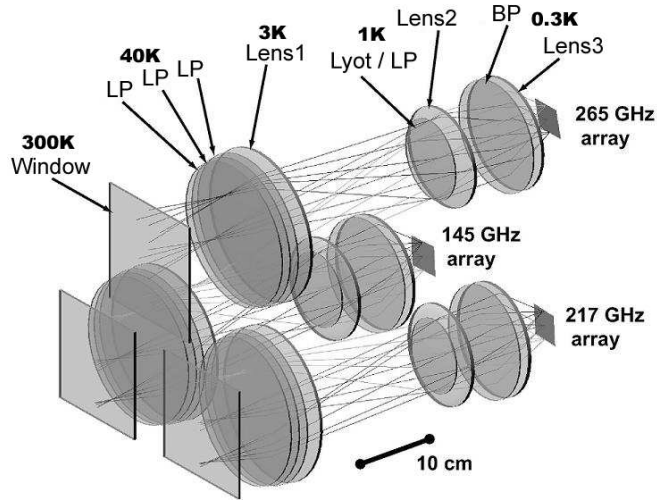}
  \caption{\label{cold_optics} 
    The cold MBAC reimaging optics.  Each frequency has a similar set
    of lenses and filters.  The 280\,GHz silicon lenses are labeled
    Lens 1 to 3 (with Lens 1 closest to the window). Infrared-blocking
    and low-pass capacitive mesh filters are all labeled LP; the
    bandpass filter is labeled BP\@. An IR-blocking filter (not shown)
    is also integrated into the window assembly.  The temperatures of
    the components decrease away from the window as indicated.  The
    bandpass filter, Lens~3, and the array are held at 0.3\,K.}
\end{figure}

A triangular configuration was chosen for the three cameras
(Figure~\ref{cold_optics}) because it packs the cameras as close as
possible to the field center, where the Gregorian image quality is
best (as measured by Strehl ratio). The close packing also maximizes
the overlap of observations.
The 280\,GHz camera is centered on the telescope's plane of symmetry
because it has the tightest diffraction requirements.  The 215\,GHz
and 145\,GHz cameras are placed symmetrically below it, allowing us to
use a single design for the two lower-frequency lens sets.  All the
lenses within each camera are parallel and share one axis.  The focal
planes and the bolometer arrays are tilted by 8\degree\ or 5\degree\
from the common axis of the lenses in their respective cameras.

\subsection{Camera components}
Figure~\ref{cold_optics} shows all three MBAC cameras.  Separate
vacuum windows are used for each camera.  The windows are made of
ultra-high molecular weight polyethylene (UHMWPE) and have
anti-reflection (AR) coatings appropriate to their respective
wavelengths.  Light entering the camera module passes through an
ambient-temperature infrared blocking ``thermal''
filter\cite{Tucker2006} (mounted just inside the vacuum window) and
three capacitive mesh filters cooled to 40\,K (marked ``LP'' in
Figure~\ref{cold_optics}); two of the three are thermal filters, and
one is a millimeter-wave low-pass filter. Together, these filters
reduce blackbody loading on the colder stages and block out-of-band
leaks in the bandpass filter (``BP''). A plano-convex silicon lens
(Lens~1) creates an image of the primary mirror near lens~2.  An
assembly holds Lens~2 and two final low-pass filters at 1~K and
contains a cold aperture stop (Lyot stop). The last two plano-convex
silicon lenses (Lenses 2 and 3) refocus the sky onto the array.  The
bandpass filter stands between these lenses, where the beam is slow
enough for the filter to be effective.  Lens~1 is cooled to 3\,K;
Lens~2 and the associated filters are cooled to 1\,K; Lens~3 and the
bandpass filter are cooled to 0.3\,K\@. The unobstructed circular
aperture of each element is large enough so that the outermost ray
that can strike any detector passes at least five wavelengths from the
aperture's edge, with the intentional exception of the Lyot stop.  The
entire camera is contained in a light-tight tube with cold black walls
to absorb stray light.  The walls are blackened with a mixture of
carbon lampblack and Stycast 2850 FT epoxy.\cite{EmersonCuming} All
walls between the bandpass filter and the array are held at the
coldest available temperature, 0.3\,K, because their emission reaches
the detectors without filtering.

Silicon was chosen as the lens material because of its high thermal
conductivity and high refractive index ($n$=3.416 at
4\,K)\cite{Parshin1995, Lamb1996}.  Pure, high-resistivity silicon
($\rho>5000\,\Omega\cdot$cm) is necessary to minimize absorption loss.
Silicon of very low electrical conductivity and low millimeter-wave
loss must be made by the float zone process rather than by the more
common (and less expensive) Czochralski process.  Float zone silicon
is available\cite{Siltronic} in diameters up to 20\,cm, restricting
our clear aperture size to 19\,cm. Alternative materials considered
for the ACT lenses included high density polyethylene (HDPE),
crystalline quartz, fused quartz, and sapphire.  Quartz and sapphire
are both more expensive to buy and more difficult to cut than silicon.
Optical designs were made using HDPE as a backup option.  However, the
plastic designs have substantially poorer image quality, a result of
making large deflections with a less refractive material.  Also, the
lower-index HDPE required much thicker lenses and consequently higher
absorption loss.

When using high refractive index materials, such as silicon,
anti-reflective coatings are critical. We have developed a method
for AR-coating silicon with quarter-wave layers of Cirlex
($n$=1.85)\cite{Lau2006ApOpt}.  Test samples show reflectivities less than
0.5\% and transmission exceeding 95\% per sample.  We expect that the
three lenses in each camera will absorb a combined 15\% of incident
light, predominantly in the Cirlex coating.  Because of the
corresponding emissivity in the lenses and their coatings, it is
necessary to cool the lenses cryogenically, reducing the power they emit.

Plano-convex lenses are used so that only one face of each lens must
be machined. The curved figures are surfaces of revolution of conic
sections plus polynomial terms in $r^4$, $r^6$, $r^8$ and $r^{10}$ to
give maximal design freedom. As the lenses were diamond turned on a
computer-controlled lathe, there was no cost penalty for adding
axially symmetric terms to the lens shapes. The curved and flat
surfaces of each lens were oriented so as to minimize
reflection-induced secondary (``ghost'') images (see
Section~\ref{ghosting}).

\subsection{Design procedure}
The Gregorian telescope design was held fixed during the cold optics
design process, while the lens shapes and positions were varied.  The
280\,GHz and 215\,GHz cameras were optimized separately.  Because the
145\,GHz and 215\,GHz cameras are placed symmetrically about the
telescope's symmetry plane---and because there is no evidence for
appreciable dispersion in silicon at millimeter wavelengths---the two
design problems are mathematically equivalent; a single camera design
was used for both.

The optimization method for the camera was similar to the method used
to design the Gregorian telescope, but with additional constraints.
Most importantly, we required a faithful image of the primary mirror
in each camera at which to place a Lyot stop.  This image quality was
quantified by tracing rays from all field points through four points
on the perimeter of the primary mirror.  The rms scatter of such ray
positions where they crossed the Lyot stop plane, projected onto the
radial direction, was included in the merit function.  Thus, an
astigmatic image of the primary elongated tangent to the stop was not
penalized, but a radial blurring was.  This additional parameter
measures the radial ray aberration at the aperture stop.

A second constraint was the effective focal length, fixed at 5.2
meters by checking the plate scale for points near the center of each
sub-field.  Finally, we found it necessary to require that the chief
ray from each field strike the focal plane at no greater than a
8\degree\ angle, which keeps the tilt of the detector plane small to
maximize absorption in the detectors.  This low-tilt requirement also
produces an approximately telecentric image, meaning that the exit
pupil is large and far from the detector plane.  A telecentric image
has the advantage that the plate scale does not depend to first order
on the relative positioning of the detector array and the lenses.

The optimizer varied up to 27 parameters in each design: three lens
positions along the optic axis, the position of the Lyot stop and its
tilt, the detector position and tilt, and the lens shapes (parameters
included curvature, conic constant, and four aspheric polynomial
terms).  We found that tilting the Lyot stop surface did not offer
enough advantage to justify the added mechanical complication and
thereafter did not allow it to tilt, reducing the number of parameters
to 25. 
The center thickness of each lens was set by requiring the edge to be
at least 2\,mm thick for mechanical strength; center thickness was not
varied by the optimizer.  We did not constrain the dimensions of the
elliptical Lyot stop.  Striking the right balance in the merit function
between optimizing the image of the sky at the detector plane and the
image of the primary at the Lyot stop was challenging.  Our most
successful approach to meeting both goals simultaneously was to make
two optimizing passes.  In the first pass, the Lyot stop image was
given large weight.  In the second, it was given zero weight, but all
parameters that affect the stop image were fixed (including shape and
placement of L1 and placement of the stop). 

The MBAC cold optics design is somewhat unusual in its use of
AR-coated silicon lenses at cryogenic temperatures.  For this reason,
we have built a prototype 145\,GHz receiver (CCam, the ``column
camera'') with a cold optics design based on the same principles as
MBAC\@.  We have tested CCam with a 1.5\,m telescope and used it
successfully to observe astronomical sources\cite{Niemack2006}, giving
us confidence in the soundness of the general design of MBAC.


\section{Design evaluation}

The full optical design was studied using both ray-tracing and
physical optics.  Most analyses were first developed for the Penn
Array Receiver at the Green Bank Radio Telescope.\cite{Dicker2005} We
present the studies most relevant to deployment, calibration, and data
analysis for ACT\@.

\begin{figure}[tbhp]
  \centering
  \includegraphics[height=12cm]{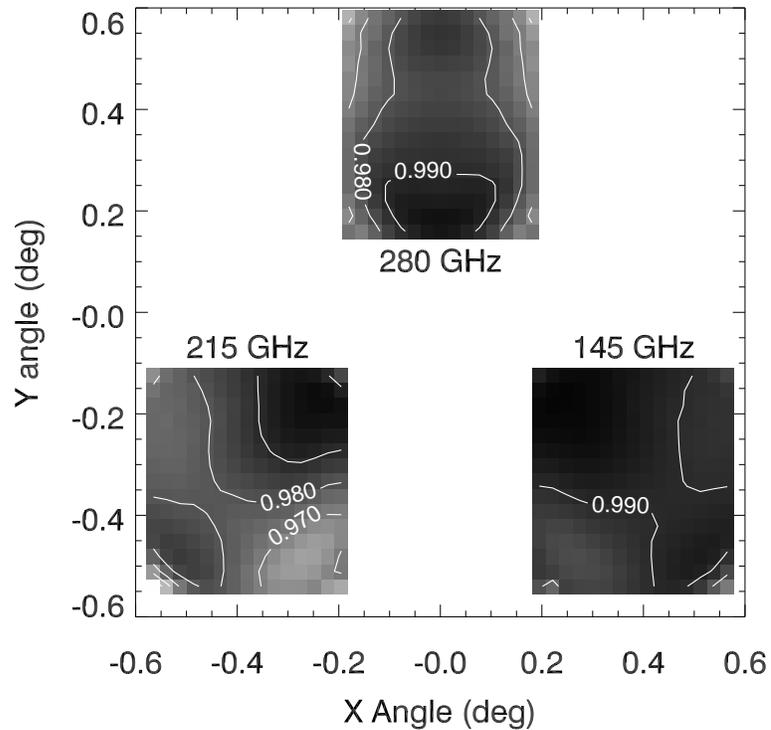}
  \caption{\label{strehls}
    Strehl ratio at points in the three ACT fields, as a function of
    field angle on the sky. The rectangular aspect of the focal plane
    array is primarily responsible for the departure from square
    fields, but anamorphic field distortion also contributes.  The
    figure also indicates the relative spacing and size of the three
    fields.  The median Strehl ratios are 0.983, 0.980, 0.991 for the
    280, 215, and 145\,GHz cameras.}
\end{figure}

\begin{figure}[th]
  \centering
  \includegraphics[height=12cm]{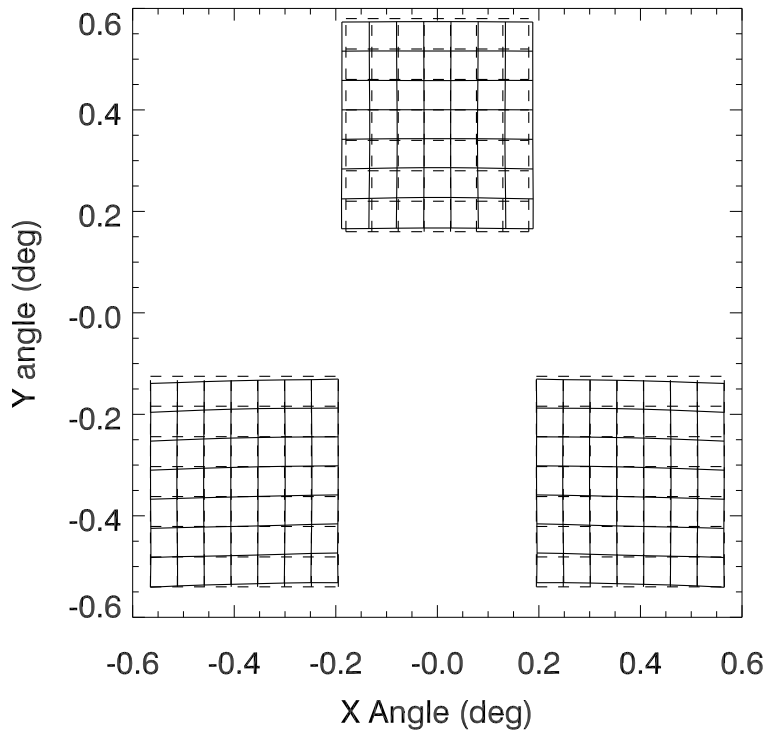}
  \caption{\label{distortion}
    Field distortion for the ACT optical design.  The dashed square
    boxes depict a notional rectangular grid of field points on the
    sky, without distortion; the solid lines indicate the image of the
    same grid after it is refocused at the detectors (assuming the
    nominal 5.20\,m effective focal length).}
\end{figure}

\subsection{Image quality}

The median Strehl ratios across the fields in the final design are
calculated to be 0.991, 0.980, and 0.983 at 145, 215, and 280\,GHz,
respectively (Figure~\ref{strehls}). The lowest Strehls corresponding
to any of the 225 field points tested in each camera are 0.971
(145\,GHz), 0.939 (215\,GHz), and 0.958 (280\,GHz).  This performance
is the baseline for comparison in the tolerance analysis
(Section~\ref{sec:tolerancing}).  These Strehl ratios establish that
all points in the field of view will be diffraction-limited.

A small amount of field distortion results from reimaging such a large
focal plane (Figure~\ref{distortion}).  One effect is anamorphic
magnification, or horizontal image stretching: the plate scale in all
cameras is $6.8'$ per cm for vertical separations, but for horizontal
separations it is only $6.4'$ per cm in the 280\,GHz camera and $6.6'$
per cm in the others.  The other effect is a shearing of the image in
the 145 and 215\,GHz cameras; lines of constant elevation are twisted
by approximately $1.4\degree$ with respect to horizontal rows of
detectors.  There is no appreciable rotation of lines of constant
azimuth.  These distortions will be taken into account in
making CMB maps from the data, but at the predicted levels, they will
not complicate our observations.


\subsection{Stop size and spillover}

The size of the elliptical Lyot stop was chosen to pass light from
only the central 97\% of the primary mirror diameter in the geometric
optics limit.  Rays were traced from many field angles through a
circular entrance pupil of diameter 291\,cm; the pupil was centered at
the primary and perpendicular to its axis.  The Lyot aperture stop in
each camera was chosen to be the largest ellipse that blocks all such
rays.  The illumination of the primary from any single field point
does not quite fill 97\% of the mirror diameter. This is because the
Lyot stop is not a perfect image of the primary for all field angles;
to make it so would degrade sky imaging performance, because the two
goals of imaging the sky and the stop compete for control over the
shape and position of the first lens.


\begin{figure}
  \centering
  \includegraphics[height=6cm]{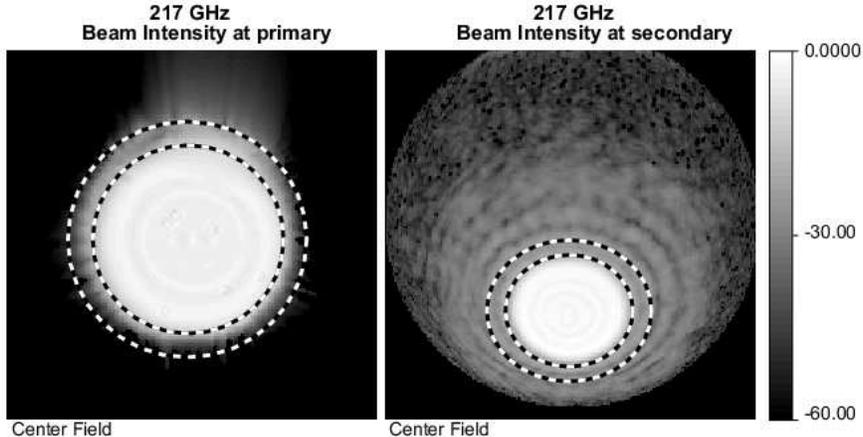}
  \caption{\label{illumination}The calculated illumination patterns of
     the primary (\emph{left}) and secondary (\emph{right}) mirrors
     projected into the $(x,y)$ plane for the central 215\,GHz
     detector.  The inner dashed circles show the edge of the mirrors
     and the outer ones show the guard ring.  Fine-scale structure in the
     right panel is an artifact of numerical precision and finite
     sampling, but all other structure is real.  (Units are dB below
     peak level.)}
\end{figure}

The Lyot stops are 43, 64, and 91 wavelengths wide at 145, 215, and
280\,GHz, respectively.  This results in a significant amount of
diffraction, so geometric analysis of the Lyot stop does not correctly
predict the spillover.  Radiation from point sources at the focal
planes was traced backwards through the ACT optics using diffraction
analysis to calculate the complex electric field at each surface.  The
intensity at each mirror surface but beyond the mirror's physical edge is
considered spillover.  A result for the primary and secondary mirrors
from a typical field is shown in Figure~\ref{illumination}.  To redirect
spillover onto the sky, each mirror is surrounded by a reflecting planar
guard ring, oriented parallel to the plane of the mirror's edge.

The illumination patterns were used to calculate the percentage of
power missing the primary and secondary mirrors.  For typical points
in the 145\,GHz field, 0.2\% of power spills over at the primary and
0.6\% at the secondary.  The values are much smaller for the 215 and
280\,GHz cameras.  The spillover would be approximately twice as large
without the guard ring.  Decreasing the size of the Lyot stop to
illuminate only 95\%, 90\% and 85\% of the main mirror diameter would
not significantly reduce spillover, because the main contribution is
from faint diffraction into wide angles at the secondary mirror.  Some
diffraction from the edge of the Lyot stop misses the secondary mirror
regardless of how much of the primary mirror is illuminated, barring
unrealistically large guard rings on the secondary.  There is therefore no
great advantage in reducing the primary mirror illumination.  We do
not expect the calculated secondary spillover to be a problem, because
the majority of it is reflected directly toward the sky.


\subsection{Tolerance analysis    
  \label{sec:tolerancing}}

Calculations were undertaken to find out how accurately optical
elements need to be placed. Position and angular displacements were
varied for all optical elements.  Other parameters varied were the
refractive indices of lenses, the shapes of surfaces, and the
temperature of the telescope.  The size of the perturbations was
increased until the rms wavefront error of a test field increased by
0.016$\lambda$\ (a reduction in the Strehl ratio of approximately
0.01).  The results of these tests were compared to the results of
dynamic finite element analysis (FEA) of the telescope structure
produced by AMEC\@.  In the following, note that a change in Strehl
produces a proportionate change in the instrument's forward gain.

The tolerancing tests were performed both with and without allowing
refocusing of the telescope using the secondary mirror (see
Section~\ref{sec:warm_opt}).  All static misalignments predicted by
FEA were easily compensated by refocusing.  We anticipate having to refocus
with each change in elevation, according to a table to be built from
beam maps made on bright, unresolved sources.
Uniform temperature changes between $-20\degree$\,C and $20\degree$\,C
do not degrade optical performance.  Temperature gradients cause
changes in the Strehl ratio well below the expected level of
variations in atmospheric transmission (and at slower timescales),
with the possible exception of direct insolation on the panels in
mid-day.  More serious are the 1\% Strehl changes expected when the
telescope is accelerated at either end of each scan.  These changes
will be scan-synchronous and might require cutting some small fraction
of the data.

Refocusing the secondary could correct for uncorrelated random
positioning errors of every optical element in a single camera if the 
misalignments do not exceed 2\,mm rms in displacement or 5\degree\ rms 
tilts. Unfortunately, though, one compromise correction must be made
for all three cameras.  Given this constraint, we find that MBAC's
optical elements need to be placed to within 1.5\,mm and 2\degree\ of
their nominal positions and orientations.  Careful mechanical design
will achieve these values.  The one dimension of substantially
tighter tolerance is the spacing between the arrays and the coldest
lenses (Lens~3).  This distance must be accurate to 0.5\,mm owing to
the fast focus onto the detectors.  Mounting the detectors and Lens~3
to a single metal structure will allow us to reach the required
accuracy.


\subsection{Ghost images
  \label{ghosting}} 

Light undergoing an even number of reflections from the nominally
transparent lens and filter surfaces in MBAC can produce secondary
(``ghost'') images.  Provided that the intensity of such images is
well below that of the diffraction-limited point spread function
(PSF), their effect can be ignored. Where this is not the case, ghost
images will act as extra sidelobes to the main beam.  We have
estimated ghosting effects using ray-tracing.  For each pair of cold
surfaces, 10,000 rays were traced through the system to the detector
focal plane, and an image was built up by adding the ray intensities
incoherently. Each ray was weighted by the reflectivity of the two
surfaces from which it reflected.  For simplicity, reflectivity was
assumed to be independent of incident angle.  Conservative (i.e.\
large) estimates were used: 3\% reflection for the window, 4\% for
each lowpass filter, 2\% for the bandpass filter, 10\% for the array,
and 1\% per surface for the antireflection-coated lenses.  To account
for diffraction, the resulting ghost images were smoothed with the PSF
of the main image.

\begin{figure}
  \centering
  \includegraphics[width=4.5in]{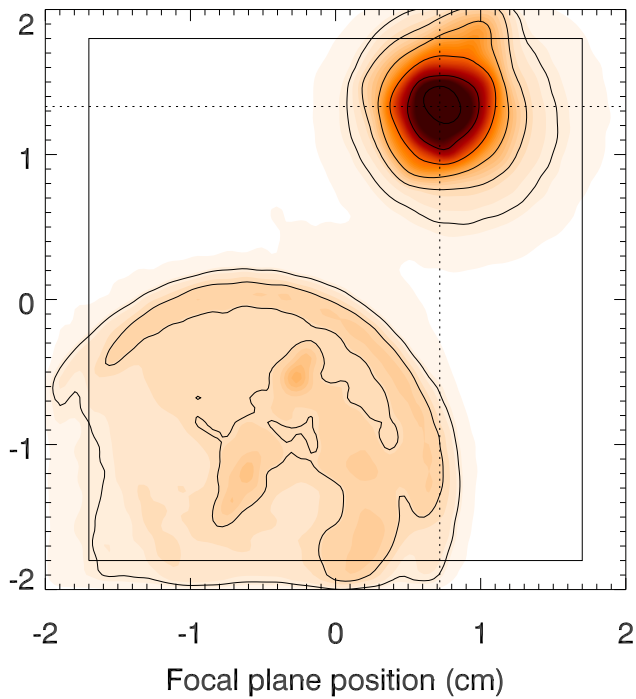}
  \caption{\label{fig:ghost}
    (Color online) 
    The secondary (ghost) images formed by stray light reflected twice
    from lens, filter, window, and array surfaces for a source imaged
    near the top of the 215\,GHz detector array.  The amplitude and
    shape of the ghost images shown here are typical of all fields,
    assuming conservative values for surface reflectivities.  The
    dotted lines cross at the center of the main image (not shown).
    The solid rectangle 3.4\,cm wide by 3.6\,cm high shows the extent
    of the focal plane array.  The shading indicates the ghost image
    intensity on a linear scale relative to the peak of the main PSF,
    saturating at $5\times10^{-4}$; the contours show intensities of
    $-47$, $-44$, $-41$, $-38$, $-35$, and $-32$\,dB relative to the
    main PSF.  The ghost intensity peaks very near the main PSF at
    approximately $-28$\,dB, where it is negligible.  The diffuse
    ghost on the opposite side of the array reaches only $-41$\,dB and
    is unlikely to affect CMB observations.}
\end{figure}

Calculations for all fields in all three cameras gave results
qualitatively similar to those shown in Figure~\ref{fig:ghost}.
Reflections between closely spaced planar elements (such as the first
three lowpass filters or the bandpass filter and the last lens) create
a ghost coincident with the main image, resulting in a peak at
approximately $-28$\,dB, centered on the main PSF\@.  This ghost is much
less intense than the diffraction-limited main image and is of little
consequence.  Light reflected from the bandpass filter and any of
several other lens or filter surfaces creates a much broader ghost
image on the opposite side of the array from the main image, with
amplitudes as high as $-41$\,dB\@.  Although this is brighter
than the diffraction sidelobes of the main PSF at that distance, the
diffuse ghost image will be below the noise of any anticipated
integration except around the brightest point sources.  With the
conservative reflectivity estimates used in the present analysis, the
ghost image's integrated intensity is approximately 2\% that of the
main PSF.  We expect a ghost image of this magnitude to be detectable
during beam-mapping studies of bright, unresolved sources (e.g.\
planets), but ghosting will not be a problem for our planned CMB
observations.


\section{Conclusions}

\begin{figure}
  \centering
  \includegraphics[height=7.5cm]{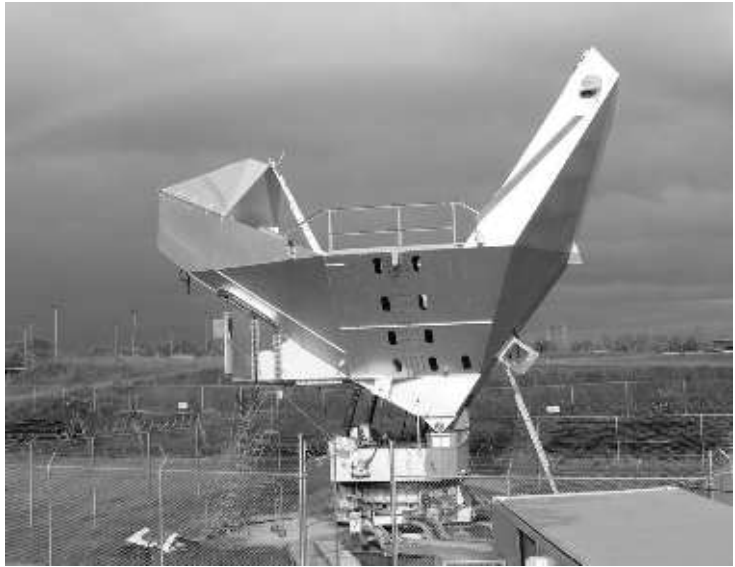}
  \caption{\label{photo} 
    The ACT telescope mostly assembled at AMEC in June 2006.  The
    inner ground screen is not completed in this photograph.}
\end{figure}

We have developed a diffraction-limited optical design that can be
used to illuminate large arrays of millimeter-wave detectors. The
design meets the requirements described in Table~\ref{requirements},
and we have studied the system properties using geometric and
diffraction analysis techniques.  The Gregorian system described in this
paper has been built in British Columbia (Figure~\ref{photo}). The
6-meter telescope is scheduled to be installed in Chile in
early 2007.  The cold optics are being built separately and will be
installed on the telescope subsequently.  A prototype camera with a
cold optics design based on silicon lenses has been built and has
observed astronomical sources using a Gregorian reflector much smaller
than ACT.

\vspace{12pt}
ACT is a dynamic collaboration, and we thank all its members for many
conversations and friendly debates.  We especially thank Bob Margolis
for his years as ACT project manager.  We wish to thank employees of
AMEC Dynamic Systems, KUKA Robotics, and Forcier Machine Design for
their countless contributions to this project.  We thank the Princeton
Physics Machine Shop for their work on the design and construction of
CCam.  We gratefully acknowledge Mandana Amiri, Bryce Burger, Rolando
Dunner, Norm Jarosik, Barth Netterfield, and Yue Zhao for thoughtful
conversations about optics, telescope operation, and related detector
matters, as well as for their collaboration in the lab and on-site at
AMEC\@.

This work was supported by the U.S. National Science Foundation
through awards AST-0408698 for the ACT project and PHY-0355328.
Princeton University and the University of Pennsylvania also provided
financial support.  Some authors received additional support from
2004-2006 SPIE Educational Scholarships in Optical Science and
Engineering (MN); an NSERC PGS-D scholarship (AH); Princeton
University Centennial and Harold Dodds Fellowships (MN and TM); and an
NSF Graduate Fellowship (TM).



\bibliographystyle{osajnl}
\bibliography{ACT_optics}

\end{document}